\documentclass[aps,twocolumn,showpacs,preprintnumbers,amsmath,amssymb,floatfix]{revtex4-2}
\usepackage{booktabs} 
\usepackage{array}     
\usepackage{graphicx}
\usepackage{longtable}
\usepackage{dcolumn}
\usepackage{bm}
\usepackage{color}
\usepackage[normalem]{ulem}
\usepackage[colorlinks=true, urlcolor=blue, linkcolor=blue, citecolor=blue]{hyperref}
\usepackage[all]{hypcap}
\newcommand{\beq}{\begin{equation}}
\newcommand{\eeq}{\end{equation}}
\newcommand{\bea}{\begin{eqnarray}}
\newcommand{\eea}{\end{eqnarray}}

\begin{document}

\title{Ab-initio calculation of magnetic exchange interactions using the spin-spiral method in VASP: Self-consistent versus magnetic force theorem approaches}

\author{Umit Daglum$^{1}$}\email[Contact email address: ]{dalumm@tcd.ie}
\author{Maria Stamenova$^{1}$}
\author{Ersoy \c{S}a\c{s}{\i}o\u{g}lu$^{2}$}
\author{Stefano Sanvito$^{1}$}
\affiliation{$^{1}$School of Physics and CRANN, Trinity College, Dublin 2, Ireland \\
$^{2}$Institute of Physics, Martin Luther University Halle-Wittenberg, 06120 Halle (Saale), Germany}

\date{\today}

\begin{abstract}
We present an \textit{ab initio} investigation of magnetic exchange interactions using the spin-spiral method implemented in the \textsc{VASP} code, 
with a comparative analysis of the self-consistent (SC) and magnetic force theorem (MFT) approaches. Using representative 3$d$ ferromagnets 
(Fe, Co, Ni) and Mn-based full Heusler compounds, we compute magnon dispersion relations directly from spin-spiral total energies and extract 
real-space Heisenberg exchange parameters via Fourier transformation. Curie temperatures are subsequently estimated within both the mean-field 
and random-phase approximations. The SC spin-spiral calculations yield exchange parameters and magnon spectra in excellent agreement with 
previous theoretical data, confirming their quantitative reliability across different classes of magnetic systems. In contrast, the MFT approach 
exhibits systematic quantitative deviations: it overestimates spin-spiral energies and exchange couplings in high-moment systems such as bcc Fe 
and the Mn-based Heuslers, while underestimating them in low-moment fcc Ni. The magnitude of these discrepancies increases strongly with magnetic 
moment size, exceeding several hundred percent in the high-moment compounds. These findings underscore the decisive role of self-consistency in 
accurately determining magnetic exchange parameters and provide practical guidance for future first-principles studies of spin interactions and 
excitations using the spin-spiral technique.
\end{abstract}

\maketitle

\section{Introduction}\label{sec1}

Understanding magnetic interactions in solids is essential for predicting and engineering the properties of a broad class 
of materials, ranging from elemental ferromagnets to complex spintronic compounds. Central to this understanding is the 
characterization of exchange interactions, which govern the collective behavior of localized and itinerant spins. These 
interactions determine key physical quantities such as the Curie temperature, magnon dispersion relations, and the stability 
of magnetic ground states. The exchange parameters \( J_{ij} \), defined within the framework of the classical Heisenberg model, provide 
a quantitative measure of the strength and range of spin–spin interactions and are widely employed in atomistic spin dynamics
simulations and multiscale modeling of magnetic phenomena~\cite{heisenberg1928theorie,antropov1996spin,ma2008large}.

First-principles calculations based on density functional theory (DFT) have become indispensable for estimating magnetic 
exchange parameters in real materials. Most computational approaches rely on the adiabatic approximation, where spin dynamics 
are assumed to be slow relative to electronic motion, allowing the total energy to be evaluated for frozen spin configurations. 
Within this framework, several methods have been developed to extract \( J_{ij} \), including the total energy mapping 
technique~\cite{archer2011exchange}, the Green’s function-based Liechtenstein–Katsnelson–Antropov–Gubanov  formalism~\cite{liechtenstein1987local}, 
and the frozen magnon or spin-spiral method~\cite{sandratskii1998noncollinear}. The latter is particularly well suited to electronic 
structure methods based on Hamiltonian diagonalization, such as the augmented spherical wave (ASW)~\cite{sandratskii2002exchange,kubler2006ab} and LMTO methods~\cite{halilov1998adiabatic}, and has also been efficiently implemented in plane-wave codes like Fleur~\cite{fleurWeb} and VASP~\cite{kresse1993ab,kresse1996efficiency}. 
These adiabatic approaches are computationally efficient and provide accurate estimates of exchange parameters and magnon 
spectra across a wide range of materials, including complex magnetic systems. However, they neglect dynamical many-body 
effects. In particular, they capture only collective magnon modes, while ignoring Stoner excitations and their coupling to 
magnons, which influence magnon lifetimes and damping. More advanced approaches, such as time-dependent DFT (TDDFT)~\cite{savrasov1998linear}
and many-body perturbation theory (MBPT) based on the T-matrix~\cite{csacsiouglu2010wannier}, go beyond the adiabatic picture and can include such effects, 
albeit at a significantly higher computational cost.

Among the various adiabatic DFT-based techniques, the spin-spiral method provides an efficient and accurate framework for describing
long-range magnetic interactions and extracting Heisenberg exchange parameters.  By exploiting the generalized Bloch theorem, 
it enables the imposition of noncollinear spin modulations, such as spin spirals, within the primitive unit cell, thereby avoiding
the computational cost of large supercells. This approach is particularly advantageous for metallic magnets with itinerant or 
weakly localized electrons, including bcc Fe and many Heusler compounds, where magnetic order emerges from extended exchange 
interactions and collective electronic effects. Overall, the spin-spiral formalism offers a powerful means of mapping the magnetic
energy landscape and characterizing complex magnetic materials.

In practical implementations, two computational approaches are commonly used for spin-spiral calculations. The first is the 
fully self-consistent (SC) method, in which the electronic structure is converged for each spiral configuration, allowing full 
relaxation of both charge and spin densities. The second is the magnetic force theorem (MFT) approach~\cite{liechtenstein1984exchange,oswald1985interaction}, 
in which the total energy is evaluated non-self-consistently by perturbing the Kohn–Sham Hamiltonian using a fixed ground-state 
charge and spin density. Although the MFT method offers a significant reduction in computational cost, its accuracy may be limited by 
the absence of SC relaxation, particularly in systems with strong hybridization or significant charge redistribution 
in spin-spiral states.

\textsc{VASP} provides native support for spin-spiral calculations based on the generalized Bloch theorem. 
In this work, we systematically compare the SC and MFT implementations 
of the spin-spiral method in \textsc{VASP}. Using a representative set of 3$d$ ferromagnets (Fe, Co, Ni) and 
Mn-based full Heusler compounds, we compute magnon dispersion relations, extract Heisenberg exchange parameters, 
and estimate Curie temperatures within both the mean-field and random-phase approximations. The SC approach yields 
exchange parameters and magnon spectra in excellent agreement with previous theoretical studies, confirming its 
quantitative reliability across different classes of magnetic \cite{halilov1998adiabatic,csacsiouglu2008role,pajda2001ab,buczek2011different,galanakis2012ab,jacobsson2017parameterisation,levzaic2013exchange}. In contrast, the MFT approach exhibits 
systematic deviations in both magnitude and trend: it overestimates spin-spiral energies and exchange couplings 
in high-moment systems such as bcc Fe and the Mn-based Heuslers, while underestimating them in low-moment 
fcc Ni. The magnitude of these discrepancies increases with magnetic moment size, reaching more than 300\% in 
the Mn-based compounds. These results highlight the decisive role of self-consistency in accurately describing 
magnetic interactions and provide practical guidance for future first-principles studies of spin excitations.

The remainder of this paper is organized as follows. Section~\ref{formalism} presents the formalism for 
calculating Heisenberg exchange parameters, magnon dispersion relations, and Curie temperatures. 
Section~\ref{comp} describes the computational setup. Section~\ref{results} contains our results and 
discussion, including a comparative analysis of spin-spiral calculations using SC and MFT 
approaches, as well as the implications of our findings. Section~\ref{conclusion} concludes the paper.

\section{Theoretical framework}\label{formalism}

The accurate determination of magnetic exchange interactions is essential for understanding and 
predicting the thermodynamic and spin-dynamical behavior of magnetic materials. In this section, 
we present the theoretical foundation for computing magnetic interactions using the spin-spiral 
method within DFT. Within the adiabatic approximation, the complex problem of 
itinerant electron magnetism can be mapped onto a classical Heisenberg model, allowing the extraction 
of exchange parameters from total energy differences between constrained spin configurations. 
We introduce the spin-spiral formalism used for this purpose and outline two complementary approaches: 
a fully SC treatment and the computationally less expensive MFT. 
We also describe how the Curie temperature can be estimated from the resulting magnon spectra.

While the formalism presented here primarily applies to single-sublattice ferromagnetic systems, in which all 
magnetic atoms occupy equivalent crystallographic sites, as in elemental 3$d$ ferromagnets (Fe, Co, Ni) and 
some Mn-based Heusler compounds, we also consider two Heusler materials with multiple magnetic sublattices. 
For these systems, we employ the generalized spin-spiral formalism for multi-sublattice magnets as developed 
in Ref.~\cite{levzaic2013exchange} to extract sublattice-resolved exchange parameters. Correspondingly, Curie 
temperatures for these compounds are estimated using the multi-sublattice MFA. 
However, our RPA analysis is restricted to single-sublattice materials, 
where the standard formalism applies. Extensions of the RPA for multi-sublattice magnets can be found 
in Refs.~\cite{csacsiog2005magnetic,rusz2005random}.

\subsection{Spin-spiral formalism}

In the adiabatic approximation, the complex problem of itinerant electron magnetism can be mapped onto a
classical Heisenberg model, which describes the magnetic behavior in terms of pairwise exchange interactions:
\begin{equation}
H = -\sum_{i \neq j} J_{ij} \, \mathbf{S}_i \cdot \mathbf{S}_j,
\end{equation}
where \( \mathbf{S}_i \) are unit vectors representing the orientation of magnetic moments at site \( i \), and \( J_{ij} \) denotes the exchange coupling between spins at sites \( i \) and \( j \).

To evaluate \( J_{ij} \) from first principles, we employ spin-spiral configurations characterized by a wavevector \( \mathbf{q} \) and a cone angle \( \theta \):
\begin{equation}
\mathbf{S}_i = \big(\sin\theta \cos(\mathbf{q} \cdot \mathbf{R}_i), \sin\theta \sin(\mathbf{q} \cdot \mathbf{R}_i), \cos\theta\big),
\end{equation}
describing a helical spin texture.

The total energy difference between a spin-spiral state and the ferromagnetic reference (\( \mathbf{q} = 0 \)) is expressed in terms of the Fourier transform of the exchange interactions:
\begin{equation}
\Delta E(\mathbf{q}, \theta) = \sin^2\theta \left[ J(\mathbf{0}) - \Re J(\mathbf{q}) \right],
\end{equation}
with
\begin{equation}
J(\mathbf{q}) = \sum_j J_{0j} e^{-i \mathbf{q} \cdot \mathbf{R}_j},
\end{equation}
where \( J(\mathbf{q}) \) is the lattice Fourier transform of the real-space exchange parameters.

The magnon dispersion relation then reads:
\begin{equation}
\omega(\mathbf{q}) = \frac{2}{M} \left[ J(\mathbf{0}) - \Re J(\mathbf{q}) \right],
\end{equation}
where \( M \) is the atomic magnetic moment (in $\mu_B$) and \( \omega(\mathbf{q}) \) is the spin-wave energy at wavevector \( \mathbf{q} \).

The inverse Fourier transform provides access to the real-space exchange couplings:
\begin{equation}
J_{0j} = \frac{1}{N \sin^2\theta} \sum_{\mathbf{q}} \Delta E(\mathbf{q}, \theta) e^{i \mathbf{q} \cdot \mathbf{R}_j},
\end{equation}
where \( N \) is the number of \( \mathbf{q} \)-points in the Brillouin zone. This method enables the extraction of Heisenberg exchange parameters directly from first-principles total energy calculations.

\subsection{Estimation of Curie temperature}

Once the exchange parameters \( J_{ij} \) are determined, \( T_c \) can be estimated using either the mean-field 
approximation (MFA) or the more accurate random-phase approximation (RPA).

In the MFA, \( T_c \) is given by:
\begin{equation}
k_BT_c^{\mathrm{MFA}} = \frac{2}{3} \sum_j J_{0j} = \frac{2}{3} J(\mathbf{0}),
\end{equation}
where \( k_B \) is Boltzmann’s constant. Although MFA tends to overestimate \( T_c \) due to its neglect of spin-wave excitations, 
it often yields reasonable results in three-dimensional systems with close-packed lattices (e.g., fcc lattice) and long-range 
exchange couplings.

A more reliable estimate is provided by the RPA, which accounts for the full magnon spectrum \cite{sandratskii2002exchange}:
\begin{equation}
k_BT_c^{\mathrm{RPA}} = \frac{2M}{3\mu_B} \left( \frac{1}{N} \sum_{\mathbf{q}} \frac{1}{\omega(\mathbf{q})} \right)^{-1},
\end{equation}
where \( \omega(\mathbf{q}) \) is the magnon energy obtained from spin-spiral calculations. The RPA captures collective spin-wave 
effects.

We note that Monte Carlo simulations provide an alternative and widely used route to estimate \( T_c \), typically yielding values 
close to those from RPA~\cite{heermann1988monte}. However, for consistency with the spin-spiral framework, we restrict our 
analysis to MFA and RPA in this work.

\subsection{Magnetic force theorem}

MFT offers a computationally inexpensive alternative to a fully SC spin-spiral 
calculations. In MFT, the spin-spiral energy is approximated by evaluating changes in the band energy using the unperturbed 
charge and spin densities of the ferromagnetic ground state. The energy change is given by:
\begin{equation}
\Delta E(\mathbf{q}, \theta) \approx  \sum_{\mathbf{k}, n} f(\epsilon_{\mathbf{k}, n}) \left[ \epsilon_{\mathbf{k}, n}(\mathbf{q},\theta ) - \epsilon_{\mathbf{k}, n}(\mathbf{0},\theta ) \right],
\end{equation}
where \( \epsilon_{\mathbf{k}, n}(\mathbf{q}) \) and \( \epsilon_{\mathbf{k}, n}(\mathbf{0}) \) are the Kohn–Sham eigenvalues for the spiral and ferromagnetic configurations, and \( f(\epsilon_{\mathbf{k}, n}) \) is the Fermi–Dirac occupation function.

This non-self-consistent treatment substantially reduces computational cost but neglects the relaxation of charge and spin densities 
in response to the noncollinear spin-spiral perturbations.

\section{Computational Details}\label{comp}

All calculations were performed using VASP, 
employing the projector augmented-wave (PAW) method~\cite{blochl1994projector}. To facilitate a meaningful comparison with available literature
results, we used different exchange-correlation functionals for different material classes. For the elemental 3$d$ ferromagnets—bcc Fe, 
fcc Co, and fcc Ni—we adopted the local density approximation (LDA)~\cite{perdew1992accurate}, which is widely used in previous studies 
for computing magnon dispersions, exchange parameters, and Curie temperatures in these systems. In contrast, for the Heusler compounds,
we employed the generalized gradient approximation (GGA) in the Perdew–Burke–Ernzerhof (PBE) formulation~\cite{perdew1996generalized}, consistent 
with the majority of prior first-principles investigations of this material class~\cite{sanvito2017accelerated}.

Experimental lattice constants were used in all calculations: 2.87~\AA{} for Fe, 3.55~\AA{} for Co, and 3.52~\AA{} for Ni.
For the Heusler compounds, we likewise used experimental lattice parameters reported in the literature~\cite{galanakis2012ab,csacsiouglu2004first}. Brillouin zone integrations were carried
out using a $\Gamma$-centered $16 \times 16 \times 16$ $k$-point mesh for SC electronic structure calculations. 
Spin-spiral calculations were performed using the generalized Bloch theorem on a uniform $10 \times 10 \times 10$ grid of spiral 
wavevectors $\mathbf{q}$, with a fixed cone angle of $\theta = 30^\circ$. All other computational parameters, including plane-wave
energy cutoff (500 eV), convergence thresholds (10$^{-6}$ eV), and smearing method, were kept consistent across both material classes to ensure comparability
of spin-spiral energies and derived magnetic properties.

In the self-consistent spin-spiral calculations, a penalty functional was employed within the constrained-spin formalism 
of \textsc{VASP} to preserve the ideal helical configuration during the self-consistency cycle. Local constraining fields, 
introduced via Lagrange multipliers, were applied to maintain the prescribed cone angle and the magnitude of the magnetic 
moments on all magnetic atoms. This approach ensures a rigid spin-spiral geometry and enables a consistent comparison of 
total energies $E(\mathbf{q})$ across different spiral vectors. Additional analyses of the penalty energy, magnetic moment 
variations, and cone-angle stability are presented in the Supplementary Material \cite{SI}.

\section{Result and Discussion}\label{results}

\subsection{Applicability of the magnetic force theorem}

MFT is widely employed to estimate magnetic excitation spectra from
a fixed ground-state electronic structure, without the need for fully SC total energy
calculations for each magnetic configuration. It has proven useful for computing spin-wave dispersions 
and extracting Heisenberg exchange parameters, particularly when computational efficiency is critical. 
However, the accuracy of the MFT can vary significantly depending on the specifics of its implementation 
and the magnetic properties of the system.

To assess the reliability of the MFT for spin-spiral calculations in \textsc{VASP}, we compared spin-spiral 
total energies obtained from MFT with those from fully SC calculations. We focus on the
elemental 3\emph{d} ferromagnets—bcc Fe, fcc Co, and fcc Ni—which serve as prototypical systems for 
itinerant magnetism and spin excitations. Our aim is to determine whether MFT provides a quantitatively 
accurate description of magnon dispersions and magnetic interactions in these materials.

\begin{figure*}[t]
\centering
\includegraphics[width=0.9\textwidth]{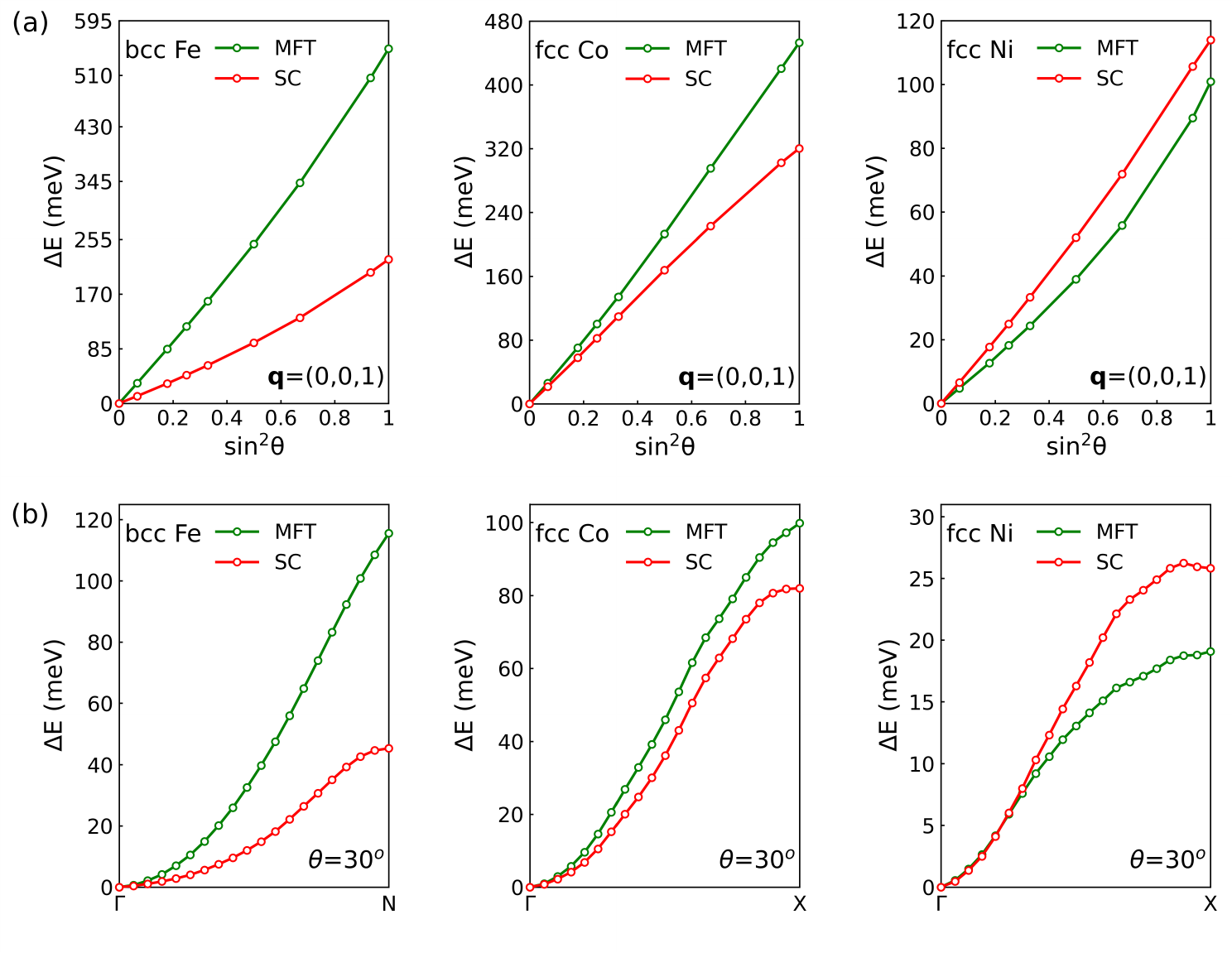}
\vspace{-0.5 cm}
\caption{Comparison of spin-spiral energies computed using the magnetic force theorem (MFT) and fully
self-consistent (SC) calculations for bcc Fe, fcc Co, and fcc Ni. (a) Spin-spiral energy as a function
of $\sin^2\theta$ at the wave vector $\mathbf{q} = (0,0,1)$, showing linear trends for both methods but 
substantial quantitative deviations, particularly for Fe. (b) Spin-spiral dispersion $E(\mathbf{q})$ 
along the high-symmetry $\Gamma$--N direction for bcc Fe and $\Gamma$--X for fcc Co and fcc Ni, calculated 
at a fixed cone angle of $\theta = 30^\circ$. The MFT results systematically deviate from the SC reference,
with overestimations in Fe and Co and underestimations in Ni near the Brillouin zone boundary.}
\label{fig1}
\end{figure*}

In the MFT approach, spin-spiral energies were evaluated non-self-consistently using a fixed ground-state 
charge density, with the magnetic structure perturbed by a transverse spiral of wave vector $\mathbf{q}$. For 
comparison, SC spin-spiral calculations were carried out with the magnetization constrained to a analogous spiral 
configuration at fixed cone angle $\theta$. This direct comparison enables us to probe the accuracy of the MFT 
across a series of ferromagnets with increasing local moment strength, from Ni to Fe.

Figure~\ref{fig1}(a) shows the spin-spiral energy as a function of $\sin^2\theta$ at $\mathbf{q} = (0,0,1)$. 
The SC results follow the expected linear dependence corresponding to the Heisenberg model across all three materials. 
While the MFT results also exhibit a linear trend, they deviate substantially in magnitude. For bcc Fe, the MFT 
overestimates the spin-spiral energy by more than a factor of two at $\sin^2\theta = 1$; in fcc Co, the 
overestimation is about 25\%; and in fcc Ni, the MFT underestimates the energy by approximately 15\%. 
These findings indicate that the MFT does not yield quantitatively reliable spin-spiral energies even for 
elemental ferromagnets, where it is most commonly applied.

This trend is further confirmed in Fig.~\ref{fig1}(b), which displays the spin-spiral dispersion $E(\mathbf{q})$ 
along the high-symmetry $\Gamma$--N direction for bcc Fe and $\Gamma$--X for fcc Co and Ni, using a fixed cone angle 
of $\theta = 30^\circ$. For Fe and Co, the MFT systematically overestimates the spin-spiral energies throughout 
the Brillouin zone, with discrepancies increasing toward the zone boundary. In Ni, MFT and SC results agree near 
$\Gamma$, but the deviation grows with increasing $|\mathbf{q}|$ and reaches roughly 30\% at the X point. 
The consistency of these deviations across all three materials suggests that they originate not from intrinsic 
magnetic behavior, but from systematic limitations of the MFT implementation in \textsc{VASP}.

These discrepancies arise from the way the magnetic force theorem is implemented in \textsc{VASP}. In non-self-consistent spin-spiral calculations, the magnetization directions are consistently rotated throughout the unit cell, including the PAW augmentation regions. However, the rotation of the exchange–correlation (XC) field cannot be uniquely assigned to individual magnetic atoms because part of the XC field resides in the interstitial region, which is shared among all atoms. This issue becomes particularly relevant in systems with multiple magnetic sublattices, where the interstitial contribution leads to ambiguities in evaluating spin-spiral energies within the MFT. Ležaic \textit{et al.}~\cite{levzaic2013exchange} demonstrated within the FLAPW framework that quantitative agreement between MFT and fully self-consistent spin-spiral calculations can be achieved only when the interstitial contribution to the XC field is neglected in the MFT energy evaluation. In contrast, when the interstitial part is included, substantial overestimations of spin-spiral energies occur. A comparable overestimation of magnon energies obtained from MFT-derived exchange parameters was recently reported by dos Santos \textit{et al.}~\cite{dossantos2025magnon}, based on a comparative study of MFT, total-energy-difference, and TDDFPT+$U$ approaches for NiO and MnO.

We therefore conclude that the MFT, in its current implementation within \textsc{VASP}, is not suitable for reliably determining 
spin-spiral energies or magnetic interaction parameters. This limitation becomes particularly critical in complex magnetic 
materials containing several magnetic atoms per unit cell, where the interstitial contributions to the exchange–correlation 
field cannot be uniquely partitioned among sublattices. In such systems, the evaluation of magnetic exchange interactions 
within the MFT framework becomes intrinsically ambiguous. In the remainder of this work, we therefore exclusively employ fully 
self-consistent spin-spiral calculations, which yield magnon dispersions and Heisenberg exchange constants in excellent agreement 
with previous theoretical studies.

\subsection{Spin-wave dispersion, exchange interactions, and Curie temperatures in Fe, Co, and Ni}

Having established the limitations of the MFT in \textsc{VASP}, we now turn
to fully SC spin-spiral  calculations to evaluate spin-wave excitations and magnetic
interactions in elemental 3\emph{d} ferromagnets. These systems, bcc Fe, fcc Co, and fcc Ni, exemplify
the progression from weak to strong ferromagnetism and from more localized to more itinerant magnetic character,
making them ideal testbeds for analyzing spin dynamics and exchange interactions beyond the MFT approximation.

Before presenting our results, it is useful to recall the two principal types of spin excitations in metallic 
ferromagnets, collective spin-wave (magnon) modes and single-particle spin-flip (Stoner) excitations, and how 
different theoretical approaches capture their distinct physical characteristics. Spin-wave excitations 
reflect the coherent precession of local magnetic moments and are often well described by effective Heisenberg 
models, in which exchange interactions are extracted either from total-energy differences or linear response. 
In contrast, Stoner excitations involve transitions between spin-split bands and are inherently incoherent, 
requiring a fully dynamical treatment of the transverse magnetic susceptibility. Such a description is only 
accessible within TDDFT or MBPT, 
which also account for magnon damping and finite lifetimes 
\cite{savrasov1998linear,csacsiouglu2010wannier}. Inelastic neutron scattering experiments reveal 
both types of excitations, although the sharp spin-wave modes dominate the low-energy region. Our SC 
spin-spiral calculations, being based on static total energies, provide access only to the spin-wave dispersion, 
without accounting for lifetime effects or damping due to coupling with Stoner excitations. As such, they do 
not capture the broadening or suppression of spin-wave modes that can occur at finite wave vectors in metallic 
systems. Nonetheless, they offer a reliable benchmark for the energy of coherent spin-wave excitations and can be 
directly compared to TDDFT or MBPT results in the low-energy regime, where damping is relatively weak.

\begin{figure}[t!]
\includegraphics[width=0.49\textwidth]{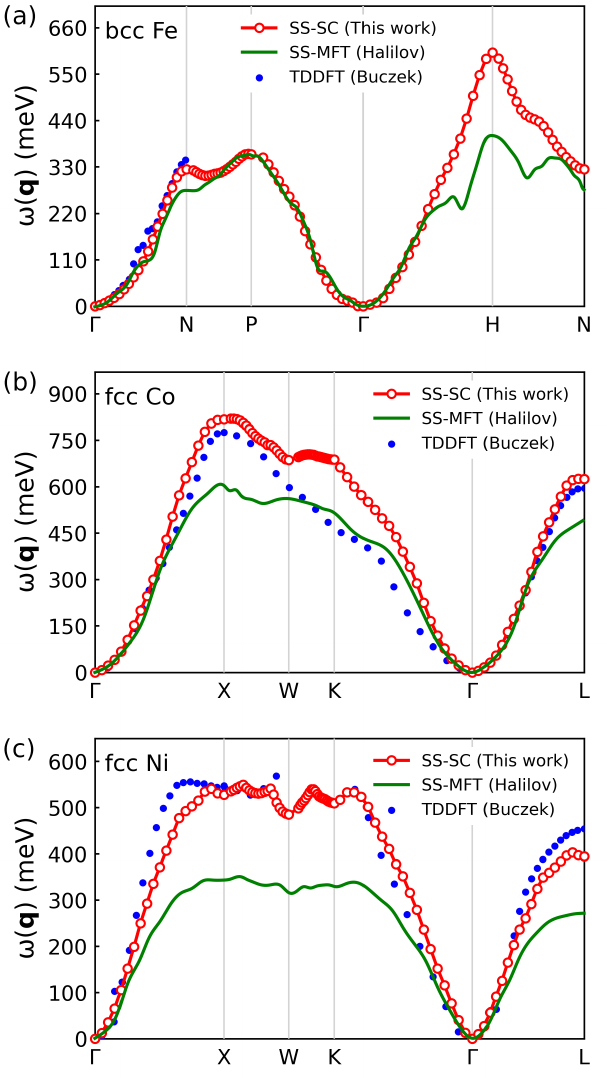}
\vspace{-0.6 cm}
\caption{Spin-wave dispersions along high-symmetry directions in the Brillouin zone for (a) bcc Fe, (b) fcc Co, and (c) fcc Ni. 
Red circles represent self-consistent spin-spiral (SS-SC) calculations from this work. These are compared with spin-spiral calculations using the magnetic force theorem (MFT) by Halilov~\textit{et al.}~\cite{halilov1998adiabatic} (solid green lines) and time-dependent density functional theory (TDDFT) results by Buczek~\textit{et al.}~\cite{buczek2011different} (blue circles).}
\label{fig2}
\end{figure}

\begin{table}[t]
\caption{Calculated and experimental magnetic properties of bcc Fe, fcc Co, and fcc Ni: lattice constants 
($a$), magnetic moments ($m$), Curie temperatures ($T_C$) from MFA and RPA, and spin-wave stiffness constants ($D$).}
\label{table1}
\begin{ruledtabular}
\begin{tabular}{lccccccc}
    & $a$ & $m$ & $T_{C}^{\mathrm{MFA}}$ & $T_C^{\mathrm{RPA}}$ &  $T_C^{\mathrm{expt.}}$  &D$_{\mathrm{calc.}}$ & D$_{\mathrm{expt.}}$     \\
    & ({\AA})   & ($\mu_B$)  & (K)  & (K)  &  (K) & (meV{\AA}$^{2}$) & (meV{\AA}$^{2}$)  \\    
\hline
bcc Fe & 2.87  & 2.25 &  1194                   &  820                     &  1043     &  220                     &  314\textsuperscript{c}  \\
       &       &      & 1077\textsuperscript{a} & 950\textsuperscript{b}   &           & 250\textsuperscript{b}   &  280\textsuperscript{d}  \\
       &       &      & 1414\textsuperscript{b} &                          &           &                          &                          \\
fcc Co & 3.55 & 1.61  & 1711                    & 1383                     & 1388      &  544                     & 510\textsuperscript{e}    \\
       &      &       & 1645\textsuperscript{b} & 1311\textsuperscript{b}  &           & 663\textsuperscript{b}   & 580\textsuperscript{f}    \\
fcc Ni & 3.52 & 0.60  & 579                     & 530                      & 627       & 757                      & 422\textsuperscript{f}    \\
       &      &       & 579\textsuperscript{a}  & 350\textsuperscript{b}  &            & 756\textsuperscript{b}   & 550\textsuperscript{g}   \\
       &      &       & 397\textsuperscript{b}  &                         &                                       &                          \\
\end{tabular}
\end{ruledtabular}
\centering
$^{a}$ Ref.~\cite{jacobsson2017parameterisation}, 
$^{b}$ Ref.~\cite{pajda2001ab}, 
$^{c}$ Ref.~\cite{stringfellow1968observation}, 
$^{d}$ Ref.~\cite{lynn1975temperature}, 
$^{e}$ Ref.~\cite{shirane1968spin}, 
$^{f}$ Ref.~\cite{pauthenet1982experimental}, 
$^{g}$ Ref.~\cite{mitchell1985low}
\end{table}

Figure~\ref{fig2} presents the spin-wave dispersions computed using the SC spin-spiral method, 
compared with MFT-based results from Halilov~\textit{et al.}~\cite{halilov1998adiabatic} and time-dependent
TDDFT results from Buczek~\textit{et al.}~\cite{buczek2011different}. Across all 
three materials, our SC dispersions closely follow the TDDFT spectra in both shape and magnitude, confirming 
the reliability of the SC spin-spiral approach for capturing low-energy spin-wave excitations. 
In contrast, the MFT results of Halilov~\textit{et al.} consistently underestimate the spin-wave energies, with the deviations becoming 
more pronounced as one moves from bcc Fe to fcc Ni. This systematic trend correlates with the decreasing 
magnetic moment and increasing itinerancy of the ferromagnetic state and reflects the known limitations of 
the MFT in itinerant systems. As emphasized by Bruno~\cite{bruno2003exchange}, the MFT becomes less accurate in materials 
with smaller local moments because it neglects the SC response of the exchange-correlation field 
to noncollinear perturbations. The agreement between SC spin-spiral and TDDFT results, despite the absence of lifetime 
effects and Stoner damping in the former, demonstrates that the static total energy approach remains quantitatively 
reliable for evaluating the energy of coherent spin waves, at least in the low- to intermediate-$\mathbf{q}$ regime 
where magnons remain well-defined.

\begin{figure}[t!]
\centering
\includegraphics[width=0.49\textwidth]{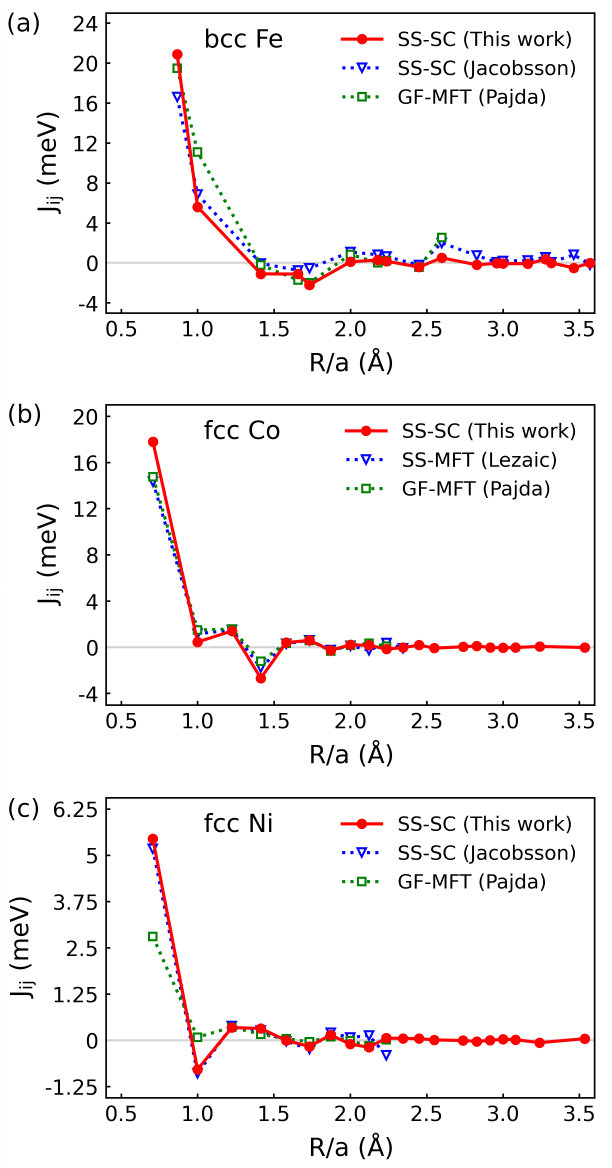}
\vspace{-0.5 cm}
\caption{Heisenberg exchange parameters $J_{ij}$ as a function of interatomic distance $R$ (in units of the lattice constant $a$)
for (a) bcc Fe, (b) fcc Co, and (c) fcc Ni. Results from our constrained self-consistent spin-spiral (SS-SC) calculations are 
compared with previous studies based on different methodologies: Pajda \textit{et al.}~\cite{pajda2001ab} (real-space MFT 
using TB-LMTO), Jacobsson \textit{et al.}~\cite{jacobsson2017parameterisation} (self-consistent spin-spiral or transverse-field method using 
the \textsc{Fleur} code), and Lezaic \textit{et al.}~\cite{levzaic2013exchange} (spin-spiral method combined with the magnetic 
force theorem in \textsc{Fleur}).}
\label{fig3}
\end{figure}

To further analyze the low-energy behavior of the computed spin-wave spectra, we extract the spin-wave stiffness 
constants $D$ by fitting the curvature of the SC dispersion near the $\Gamma$ point. These values, presented in 
Table~\ref{table1}, reflect the increasing rigidity of the spin system from Fe to Ni, with $D$ rising from 220~meV{\AA}$^2$ 
in bcc Fe to 544~meV{\AA}$^2$ in fcc Co and reaching 757~meV{\AA}$^2$ in fcc Ni. The results for Fe and Co are in 
good agreement with experimental data and prior theoretical estimates, demonstrating the accuracy of the SC 
spin-spiral approach in itinerant ferromagnets. In fcc Ni, however, the calculated $D$ substantially overshoots
the experimental range. This discrepancy highlights the breakdown of the Heisenberg mapping in the weak-moment,
strongly itinerant regime, where magnon–Stoner coupling, hybridization effects, and additional optical magnon 
branches—observed in experiments and captured by more advanced frameworks such as MBPT~\cite{csacsiouglu2010wannier}—
become increasingly important. While such effects lie beyond the scope of static total-energy calculations, the 
extracted stiffness constants remain quantitatively accurate in the long-wavelength limit.

While the spin-wave stiffness constant $D$ provides a useful measure of the overall rigidity of the magnetic
system in the long-wavelength limit, a more detailed understanding of the microscopic magnetic interactions
requires access to the underlying real-space exchange couplings. To this end, we extract the Heisenberg exchange 
parameters $J_{ij}$ by Fourier transforming the total energies obtained from SC spin-spiral 
calculations. Since the spin-spiral method operates in reciprocal space, the spatial resolution and maximum
interaction range of the $J_{ij}$ parameters are determined by the density of the $\mathbf{q}$-point mesh used
in the Brillouin zone sampling. In our case, the chosen $\mathbf{q}$-mesh allows us to resolve exchange 
interactions up to a distance of $3.5a$, where $a$ is the lattice constant of the respective material. This 
cutoff captures both the dominant near-neighbor couplings and longer-range contributions that are particularly
important in metallic systems. The resulting exchange profiles enable a direct comparison with earlier studies 
employing alternative methodologies.

Figure~\ref{fig3} presents the calculated Heisenberg exchange parameters $J_{ij}$ for bcc Fe, fcc Co, 
and fcc Ni as a function of interatomic distance $R$ (expressed in units of the lattice constant $a$). 
In all three materials, the dominant contribution arises from the nearest-neighbor (NN) interaction, 
which reaches approximately 21~meV in Fe, 18~meV in Co, and 6~meV in Ni. These large NN values can be 
attributed to the strong overlap of partially filled 3$d$ orbitals in these elemental ferromagnets. The 
second-nearest-neighbor (2NN) couplings are significantly weaker—about 5~meV for Fe, nearly vanishing for 
Co, and close to $-1$~meV for Ni. For larger interatomic separations, the $J_{ij}$ values become small but 
exhibit oscillatory behavior, reflecting the itinerant nature of magnetism and the metallic character of
the materials.

A noticeable difference appears in the spatial profile of the exchange couplings among the three systems. 
In bcc Fe, the $J_{ij}$ parameters extend over longer distances and exhibit more pronounced oscillations, 
which is consistent with its classification as a weak ferromagnet. In such systems, majority-spin 3$d$ 
states partially cross the Fermi level, enabling magnetic interactions to be mediated by itinerant $s$ 
electrons and partially delocalized $3d$ states through an RKKY-like mechanism. In contrast, fcc Co and especially 
fcc Ni exhibit stronger ferromagnetism, with fully occupied majority-spin $3d$ bands and a lower density 
of states at the Fermi level, resulting in more rapidly decaying and short-ranged exchange interactions.

Compared to the real-space MFT calculations of Pajda~\textit{et al.}~\cite{pajda2001ab}, our self-consistent 
spin-spiral (SS-SC) results systematically yield larger $J_{ij}$ values. This is consistent with the known 
underestimation of magnetic excitations by the MFT approach, particularly in systems with small magnetic moments. The discrepancy is especially pronounced in fcc Ni, where exchange interactions are highly
sensitive to the SC treatment of the exchange-correlation field. Our $J_{ij}$ profiles for Fe 
and Ni agree very well with those of Jacobsson~\textit{et al.}~\cite{jacobsson2017parameterisation}, who employed a SC
spin-spiral method using the \textsc{Fleur} code. For Co, our results are in good agreement with the spin-spiral
MFT data of Lezaic~\textit{et al.}~\cite{levzaic2013exchange}, which are also consistent with the findings of Pajda. 
These comparisons support the reliability and transferability of our reciprocal-space spin-spiral approach for 
extracting real-space exchange parameters in 3$d$ ferromagnets.

To further validate our approach, we benchmark our results against spin-spiral calculations in the 
all-electron \textsc{Fleur} code. For bcc Fe and fcc Ni, our $J_{ij}$ values show excellent agreement 
with the SC spin-spiral results of Jacobsson~\textit{et al.}~\cite{jacobsson2017parameterisation}, particularly
in Ni where both the magnitude and decay profile are nearly identical. For fcc Co, we compare to the 
MFT-based spin-spiral results of Ležaic~\cite{levzaic2013exchange}, which yield a similar trend but systematically
lower magnitudes, again reflecting the underestimation intrinsic to the MFT approximation (see Fig.\ref{fig3}). Notably, the 
Ležaic and Pajda data for Co align closely with each other, reinforcing the consistency between real-space 
and reciprocal-space MFT implementations. These comparisons confirm that our SC spin-spiral
approach captures both the magnitude and spatial decay of the exchange parameters with high accuracy, making
it a reliable method for studying magnetic interactions in itinerant ferromagnets.

Finally, we estimate the Curie temperatures ($T_C$) of bcc Fe, fcc Co, and fcc Ni using both the MFA and the more accurate RPA, based on the extracted Heisenberg 
exchange parameters. The results are listed in Table~\ref{table1}, together with experimental values and 
previous theoretical estimates. As expected, MFA systematically overestimates $T_C$ due to its neglect of 
spin-wave fluctuations, while RPA generally provides improved agreement with experiment. For fcc Co and Ni, 
the RPA results are in good agreement with experimental values: 1383~K versus 1388~K for Co, and 530~K versus 
627~K for Ni, significantly outperforming the corresponding MFA estimates. For bcc Fe, however, the RPA value 
of 820~K underestimates the experimental $T_C$ of 1043~K and is the lowest among all available theoretical 
predictions, including prior RPA-based spin-spiral estimates in the range of 950–1000~K~\cite{pajda2001ab}.

\subsection{Exchange interactions and Curie temperatures in Mn-based full Heusler compounds}

Heusler compounds offer a versatile platform for exploring diverse magnetic exchange mechanisms due 
to their rich chemical tunability, multiple magnetic sublattices, and sensitivity to structural and 
electronic parameters~\cite{graf2011simple}. Depending on their composition and the number 
of magnetic atoms per unit cell, these materials can host a wide range of magnetic phases, including 
ferromagnetic, ferrimagnetic, and noncollinear orders. In compounds with multiple magnetic sublattices, 
such as Co$_2$MnSi, both intra- and inter-sublattice couplings shape the magnetic ground state~\cite{csacsiouglu2005exchange}. 
By contrast, many Mn-based full Heuslers contain only a single magnetic atom per unit cell, 
with magnetic exchange typically dominated by long-range, indirect interactions mediated by conduction electrons.

Given the diversity of exchange mechanisms in these systems, we have also examined the applicability 
of the MFT to Mn-based full Heusler compounds. As detailed in the Supplementary 
Material \cite{SI}, a direct comparison between SC and MFT spin-spiral energies reveals that, 
in multisublattice systems such as Pd$_2$MnSn, the MFT within \textsc{VASP} substantially overestimates 
the spin-spiral energies—by up to approximately 380\,\% in our tests. This finding demonstrates that 
the current MFT implementation in \textsc{VASP} cannot be reliably applied to complex magnetic structures 
containing several magnetic atoms per unit cell. Consequently, all Heisenberg exchange parameters 
reported below are obtained from fully SC spin-spiral calculations.

Having established the reliability of the SC approach, we now assess its predictive 
accuracy for four representative Mn-based full Heusler compounds: Cu$_2$MnAl, Ni$_2$MnSn, Pd$_2$MnSn, 
and Ni$_2$MnGa. These systems serve as a valuable benchmark, owing to the availability of reliable 
experimental data and detailed theoretical studies, particularly those of \c{S}a\c{s}{\i}o\u{g}lu 
\textit{et al.} \cite{csacsiouglu2008role}. Before analyzing their exchange behavior, it is helpful to recall 
the trends observed in elemental 3\textit{d} ferromagnets (bcc Fe, fcc Co, and fcc Ni), discussed in the 
previous subsection. In these systems, magnetic exchange is dominated by strong nearest-neighbor couplings 
arising from direct 3\textit{d}--3\textit{d} orbital overlap. The resulting \( J_{ij} \) values are large 
at short distances, while further-neighbor interactions are significantly weaker and often oscillatory 
due to indirect mediation by conduction (\textit{sp}) electrons.

The four Mn-based Heusler compounds studied here share a common structural feature: a single Mn atom per 
unit cell and Mn--Mn distances exceeding 4\,\AA{}. These large separations result in negligible direct 
3\textit{d}--3\textit{d} orbital overlap, rendering direct Mn--Mn exchange inefficient. As a consequence, 
magnetic interactions are primarily mediated by conduction electrons through indirect mechanisms such as 
RKKY-type exchange or antiferromagnetic superexchange. Among the compounds considered, Cu$_2$MnAl exhibits 
the strongest nearest-neighbor Mn--Mn exchange interaction ($\sim$11\,meV) and a sizable second-nearest-neighbor 
coupling ($\sim$5\,meV), with subsequent couplings showing an oscillatory decay.

\begin{table}[t]

\caption{\label{table2} 
Lattice constants ($a$), atom-resolved spin magnetic moments ($m_\text{X}$ and $m_\text{Y}$), total magnetic moments, and Curie temperatures for the studied Mn-based full Heusler compounds. Both mean-field (MFA) and random phase approximation (RPA) estimates are reported alongside available experimental data. For comparison, selected results from previous theoretical studies are also included.}
\begin{ruledtabular}
\begin{tabular}{llllllll}
Material & $a$  & $m_{\text{X}}$ & $m_{\text{Y}}$ & m$_\text{total}$& $T_c^{\text{MFA}}$  & $T_c^{\text{RPA}}$ & $T_c^{\text{Exp}}$  \\
X$(_{2})$YZ   &({\AA}) &  ($\mu_B$) &  ($\mu_B$) & ($\mu_B$) & (K)  & (K) & (K)   \\ 
\hline
Cu$_{2}$MnAl & 5.95 & 0.075                   & 3.55                     & 3.56                    & 1008                    & 713                      & 603\textsuperscript{a}     \\ 
             &      & 0.02\textsuperscript{b} & 3.67\textsuperscript{b} & 3.60\textsuperscript{b} & 970\textsuperscript{b} & 635\textsuperscript{b}   & \\
Pd$_{2}$MnSn & 6.38 & 0.07                    & 3.97                     & 4.11                    & 275                     & 242                      & 189\textsuperscript{c} \\ 
             &      & 0.07\textsuperscript{b} & 4.08\textsuperscript{b}  & 4.16\textsuperscript{b} & 252\textsuperscript{b}  & 178\textsuperscript{b}   & \\
Ni$_{2}$MnGa & 5.85 & 0.33                    & 3.44                     & 4.03                    & 402                     & -                        &380\textsuperscript{d} \\
             &      & 0.29\textsuperscript{e} & 3.57\textsuperscript{e}  & 4.09\textsuperscript{e} & 389\textsuperscript{e}  &                           &  \\
Ni$_{2}$MnSn & 5.99 & 0.23                    & 3.53                     & 3.92                    & 397                     & -                         & 360\textsuperscript{d} \\ 
             &      & 0.21\textsuperscript{e} & 3.72\textsuperscript{e}  & 4.08\textsuperscript{e} & 358\textsuperscript{e}  &                           &                        \\
\end{tabular}
\end{ruledtabular}
\begin{tabbing}
$^{a}$Ref.\,\cite{noda1976spin} $^{b}$Ref.\,\cite{galanakis2012ab} $^{c}$Ref.\,\cite{tajima1977spin}  $^{d}$Ref.\,\cite{webster1988} $^{e}$Ref.\,\cite{csacsiouglu2004first} \
\end{tabbing}
\end{table}

\begin{figure*}[t]
\centering
\includegraphics[width=0.97\textwidth]{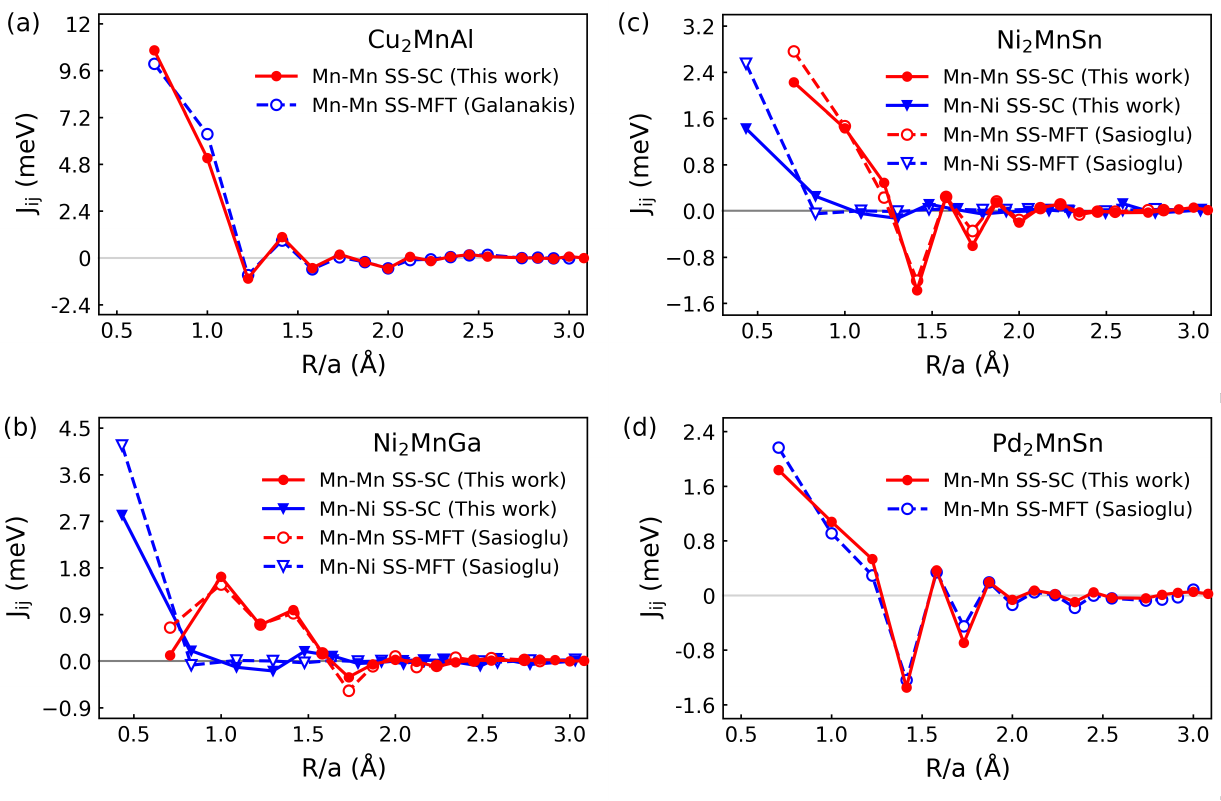}
\vspace{-0.3 cm}
\caption{Heisenberg exchange parameters $J_{ij}$ as a function of interatomic distance $R$ (in units of the lattice constant $a$) for the Mn-based full Heusler compounds: (a) Cu$_2$MnAl, (b) Ni$_2$MnGa, (c) Ni$_2$MnSn, and (d) Pd$_2$MnSn. For Cu$_2$MnAl and Pd$_2$MnSn, only Mn--Mn exchange interactions are shown and compared with the results of Galanakis \textit{et al.}~\cite{galanakis2012ab}. In the Ni-based compounds Ni$_2$MnGa and Ni$_2$MnSn, both Mn--Mn and Mn--Ni exchange parameters are displayed, with literature values taken from the multi-sublattice study of Şaşıoğlu \textit{et al.}~\cite{csacsiouglu2004first}. All exchange interactions were obtained from our constrained self-consistent spin-spiral (SS-SC) calculations.}
\label{fig4}
\end{figure*}

In Ni$_2$MnGa, the first three Mn--Mn interactions are ferromagnetic, although the nearest-neighbor term 
is nearly negligible. This compound also features a strong nearest-neighbor Mn--Ni exchange of $\sim$4\,meV, 
enabled by short Mn--Ni separation and 3\textit{d}--3\textit{d} hybridization, despite the small Ni magnetic 
moment ($\sim$0.3\,$\mu_\mathrm{B}$). Ni$_2$MnSn and Pd$_2$MnSn, being iso-electronic and structurally 
similar, exhibit nearly identical $J_{ij}$ profiles: the first three Mn--Mn exchange parameters are 
ferromagnetic, as expected from the empirical Castelliz-Kanomata argument~\cite{Castelliz1955,Kanomata1987},
followed by long-range oscillatory couplings. Minor differences stem from variations in 
their lattice constants (see Table~\ref{table2}). In Ni$_2$MnSn, a nearest-neighbor Mn--Ni exchange 
interaction is also observed, though it is smaller in magnitude than the corresponding Mn--Mn interaction.

Across all four compounds, the calculated exchange parameters show excellent agreement with previous 
MFT-based results, particularly those reported by Galanakis \textit{et al.}~\cite{galanakis2012ab} and 
\c{S}a\c{s}{\i}o\u{g}lu \textit{et al.}~\cite{csacsiouglu2004first}, thereby validating the accuracy of our 
SC spin-spiral approach. It is worth noting that the large magnetic moments characteristic 
of Mn-based Heuslers lead to negligible differences between MFT and fully SC spin-spiral 
calculations, as also confirmed in Ref.~\cite{csacsiouglu2004first}. This agreement forms a solid foundation 
for evaluating the magnetic ordering tendencies and Curie temperatures in the subsequent analysis.

The Curie temperatures computed from the extracted Heisenberg exchange parameters are summarized 
in Table~\ref{table2}, alongside experimental values and results from previous theoretical studies. 
For all four compounds, we report the MFA estimates of $T_c$, which provide a convenient upper 
bound but are known to overestimate the transition temperature due to their neglect of collective spin 
fluctuations. To obtain more realistic predictions, we also evaluate $T_c$ within the RPA, which incorporates spin-wave excitations. However, our RPA implementation is 
currently limited to systems with a single magnetic sublattice, and hence RPA estimates are only 
provided for Cu$_2$MnAl and Pd$_2$MnSn. For these two compounds, the RPA $T_c$ values are in excellent 
agreement with previous MFT-based studies and show a marked improvement over MFA in terms of quantitative 
agreement with experimental data. Specifically, for Cu$_2$MnAl we find $T_c^\mathrm{RPA} = 713$\,K, 
close to the experimental value of 603\,K, while Pd$_2$MnSn yields $T_c^\mathrm{RPA} = 242$\,K compared to the experimental 189\,K. In the case of Ni$_2$MnSn and Ni$_2$MnGa, which contain both Mn and Ni magnetic 
atoms, RPA calculations would require a full multi-sublattice formalism. We therefore restrict our 
analysis to the MFA values for these compounds, which are found to be in good agreement with previous 
theoretical work and within 10\% of the experimental values. Overall, the computed $T_c$ values 
corroborate the trends observed in the exchange interactions and further confirm the reliability of
the SC spin-spiral approach for estimating finite-temperature magnetic properties.

\section{Summary and Conclusions}\label{conclusion}

In this work, we have carried out a comprehensive first-principles study of magnetic exchange interactions using the spin-spiral method implemented in the \textsc{VASP} code, comparing the fully self-consistent (SC) approach with the magnetic force theorem (MFT) variant. Our analysis covered two classes of systems: elemental 3\textit{d} ferromagnets (bcc Fe, fcc Co, and fcc Ni) and representative Mn-based full Heusler compounds (Cu$_2$MnAl, Ni$_2$MnSn, Pd$_2$MnSn, and Ni$_2$MnGa). For each material, we computed spin-spiral total energies, extracted real-space Heisenberg exchange parameters via Fourier transformation, and estimated Curie temperatures using both the mean-field approximation (MFA) and the random-phase approximation (RPA).

Our results show that SC spin-spiral calculations yield magnon dispersions and exchange parameters in excellent agreement with previous theoretical data, confirming their reliability across diverse magnetic systems. In contrast, the MFT approach exhibits systematic quantitative deviations: it overestimates spin-spiral energies and exchange couplings in bcc Fe and fcc Co, while underestimating them in fcc Ni. The magnitude of these deviations increases markedly with magnetic moment size and degree of localization. In Mn-based full Heusler compounds with large Mn moments of about 4\,$\mu_\mathrm{B}$, the MFT errors become particularly pronounced, exceeding several hundred percent relative to the self-consistent results. These findings demonstrate that the current MFT implementation in \textsc{VASP} does not provide a quantitatively reliable description of magnetic interactions, especially in materials with large local moments or multiple magnetic sublattices.

Overall, our study underscores the necessity of fully self-consistent spin-spiral calculations for obtaining accurate exchange parameters and spin-wave spectra within \textsc{VASP}. The observed deviations—ranging from moderate underestimations in low-moment systems to strong overestimations in high-moment magnets—highlight the quantitative limitations of the MFT and establish self-consistency as a prerequisite for reliable predictions. These results provide a solid methodological benchmark for future first-principles investigations of magnetic ordering and spin excitations in complex materials.

\begin{acknowledgments}
This work was supported by Science Foundation Ireland under Grant No.~18/SIRG/5515. The authors acknowledge the DJEI/DES/SFI/HEA, Trinity College Dublin Research IT and the Irish Center for High-End Computing (ICHEC) for providing computational resources. U.D. gratefully acknowledges Martin Schlipf for insightful discussions and valuable feedback. E.\c{S}. acknowledges financial support from the Deutsche Forschungsgemeinschaft (DFG) through the Collaborative Research Center CRC/TRR 227.

\end{acknowledgments}

\section*{Data Availability Statement}

Data available on request from the authors

\nocite{*}
\bibliography{main_ref}

\end{document}